\documentclass[aip,jap,reprint,floatfix]{revtex4-1}

\usepackage{graphicx}
\usepackage[squaren,thinspace,textstyle]{SIunits}
\addunit{\evolt}{e\volt}

\newcommand{\iside}{\ensuremath{I_{\text{side}}{}}}
\newcommand{\iC}{\ensuremath{I_{\text{C}}}}
\newcommand{\iE}{\ensuremath{I_{\text{E}}}}
\newcommand{\eF}{\ensuremath{E_{\text{F}}}}
\newcommand{\vBC}{\ensuremath{V_{\text{BC}}}}
\newcommand{\eBC}{\ensuremath{E_{\text{BC}}}}
\newcommand{\vBE}{\ensuremath{V_{\text{BE}}}}
\newcommand{\vE}{\ensuremath{V_{\text{E}}}}
\newcommand{\vEt}{\ensuremath{V_{\text{E}}^{\text{th}}}}
\newcommand{\lee}{\ensuremath{l_{\text{ee}}}}

\newcommand{\ekin}{\ensuremath{E_{\text{kin}}}}

\usepackage{color}
\newcommand{\todo}[1]{\textcolor{blue}{ TODO #1}}

\begin{document}

\title{An electron jet pump: The Venturi effect of a Fermi liquid}



\author{D. Taubert}
\affiliation{Center for NanoScience and Fakult\"at f\"ur Physik,
Ludwig-Maximilians-Universit\"at,
Geschwister-Scholl-Platz 1, 80539 M\"unchen, Germany}

\author{G. J. Schinner}
\affiliation{Center for NanoScience and Fakult\"at f\"ur Physik,
Ludwig-Maximilians-Universit\"at,
Geschwister-Scholl-Platz 1, 80539 M\"unchen, Germany}

\author{C. Tomaras}
\affiliation{Center for NanoScience and Fakult\"at f\"ur Physik,
Ludwig-Maximilians-Universit\"at,
Geschwister-Scholl-Platz 1, 80539 M\"unchen, Germany}

\author{H. P. Tranitz}
\affiliation{Institut f\"ur Experimentelle Physik,Universit\"at Regensburg,
93040 Regensburg, Germany}

\author{W. Wegscheider}
\affiliation{Solid State Physics Laboratory, ETH Zurich, 8093 Zurich,
Switzerland}

\author{S. Ludwig}
\affiliation{Center for NanoScience and Fakult\"at f\"ur Physik,
Ludwig-Maximilians-Universit\"at,
Geschwister-Scholl-Platz 1, 80539 M\"unchen, Germany}

\date{\today}

\begin{abstract}
A three-terminal device based upon a two-dimensional electron system is investigated in the regime of non-equilibrium
transport. Excited electrons scatter with the cold Fermi sea and transfer energy and momentum to other electrons. A
geometry analogous to a water jet pump is used to create a jet pump for electrons. Because of its phenomenological
similarity we name the observed behavior the ``electronic Venturi effect''.
\end{abstract}


\maketitle

\section{introduction}

The Venturi effect in hydrodynamics describes the relation between the pressure of an inviscid fluid and the
cross-section of the tubing it flows through, as a reduced cross-section leads to
reduced pressure. One of the more famous applications of this phenomenon is the
water jet pump indroduced by Bunsen in 1869 \cite{bunsen} in which the decrease
of fluid pressure in a constriction is used for evacuating a side port.
Beyond the bottleneck, the fluid reaches a wider collector tube and decelerates.
Here we present a similar system, an ``electron jet pump,'' built from a
degenerate two-dimensional electron system, a Fermi liquid. ``Hydrodynamic''
effects in Fermi liquids have been studied
theoretically \cite{sasha} and experimentally\cite{dyakonov}, however, ``hydrodynamic'' has been used in
different ways.
While e.\ g., Ref.\ \onlinecite{dyakonov} describes a system governed by a set of equations essentially identical to
those
describing hydrodynamics and Ref.\ \onlinecite{gardner} extends these equations to a quantum-mechanical regime, Ref.\
\onlinecite{sasha} along with the experiments
presented here use hydrodynamics as a qualitative analogy since the results are very similar from a phenomenological
point of view. The electronic analogy of the Venturi effect has
been introduced in Ref.\ \onlinecite{hotelectrons}; other experiments describing related physics but, in part, based
upon
different effects have been performed since the 1990s.\cite{brill,kaya}

\section{device and setup}
\begin{figure}
\includegraphics[width=\columnwidth]{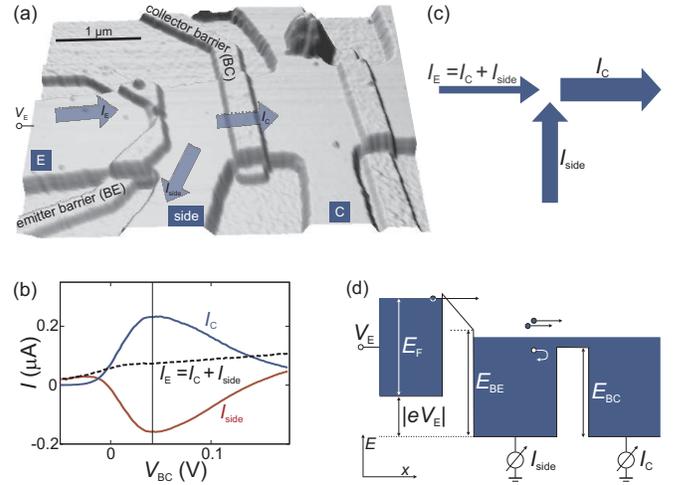}
\caption{\label{fig-sample}(Color online) (a) Atomic force micrograph of the sample. Elevated areas represent
metal gates fabricated on top of a hall bar defined by wet etching.
Definition of positive current directions (direction of electron
flow) is marked by arrows. (b) Three
currents defined in (a) as a function of voltage applied to gate BC for $\vBE
=\unit{-0.925}{\volt}$, $\vE = \unit{-155.3}{\milli\volt}$ (c) Diagram of arrows showing the actual current
directions, at position marked in (b) by a vertical line, with the arrow width resembling the magnitude of current, and
(d) model of electronic Venturi effect (see main text).
}
\end{figure}

Fig.\ \ref{fig-sample}(a)
 shows an atomic force micrograph of the device used to
demonstrate the electronic Venturi
effect. It has been fabricated from a GaAs/AlGaAs heterostructure containing a
two-dimensional electron system (2DES) \unit{90}{\nano\meter} below the
surface. The 2DES has a mobility of $\mu=1.4 \times 10^6\,\text{cm}^2/\text{Vs}$
(at $ T \approx 1\,\text{K}$) and a Fermi
energy of $\eF = 9.7\,\text{meV}$ (carrier density $n_s=2.7 \times
10^{15}\,\text{m}^{-2}$). The elastic mean-free path 
$l_{\text{m}}\simeq 12\; \mu\text{m}$ is much larger than the
sample dimensions. All measurements presented here have been performed in a
$^3$He cryostat at a bath temperature of \unit{260}{\milli\kelvin}, but similar
results have been obtained in a temperature range of $20\,\text{mK}\le
T_{\text{bath}}\le 20\,$K in several comparable samples.

A hall-bar-like structure created by wet etching defines the general layout of the device with a central area with
several terminals connected to ohmic contacts (not visible). Three of them
are used in the experiments shown here, namely the emitter ``E'', ``side'' contact,
and collector ``C''. Additionally, metallic gates [elevated in Fig.\
\ref{fig-sample}(a)] are used to electrostatically define the barriers. A quantum
point contact, called the ``BE'' (emitter
barrier), and a broad collector barrier ``BC'' are used for demonstrating the electronic Venturi effect; the
device contains more gates, though. All measurements presented here have been performed with the QPC as emitter, but 
using a broad barrier as ``BE'' produces very similar results. The special nature of a QPC is, therefore, not crucial.
The
terminal in the top right corner of Fig.\ \ref{fig-sample}(a) did not carry
current, which might be related to the contamination visible in the micrograph.

\section{electron jet pump}
\label{sec-jetpump}

A bias voltage, \vE, is applied to the emitter contact
while ``side'' and ``C'' are grounded via low-noise current amplifiers. At the emitter, a current, \iE, flows which we
define to be positive if electrons are injected into the device ($\vE < 0$). In
a network of ohmic resistors, the electrons would be expected to
leave the device at the two contacts ``side'' and ``C''; we thus define the
resulting currents, \iside\ and \iC, to be positive in such an
ohmic situation. For the definitions applied here, Kirchhoff's current law
therefore reads, $\iE = \iC + \iside$ [also compare arrows in Fig.\ \ref{fig-sample}(a)].

Fig.\ \ref{fig-sample}(b) shows the simultaneously measured dc currents, \iC\ and
\iside, along with the derived quantity, \iE, as a function of \vBC, which is the voltage
applied to the collector barrier. In most of the plot, non-ohmic behavior is
observed as \iC\ exceeds \iE, equivalent to a negative side current. This
behavior is visualized in Fig.\ \ref{fig-sample}(c) which shows three arrows resembling the currents for a situation
marked in Fig.\ \ref{fig-sample}(b) by a vertical line. The width of the arrows stands for the
magnitude of the
respective currents. As more electrons leave the device at ``C'' than are injected at ``E'', this effect can be viewed
as amplification of the injected current. Alternatively, and concurrent with the hydrodynamic analogy, it can be
interpreted as jet pump behavior, as electrons are drawn into the device at the side port. 

The observed effect can be understood as follows. Due to the voltage drop of \vE\ across the emitter barrier BE, which
is close to pinch-off, electrons are injected into the central region of the device with a kinetic energy of
approximately $\left| e \vE + \eF\right|$, which is \unit{163}{\milli\evolt} in the case of Fig.\
\ref{fig-sample}(b). Electrons with such an energy scatter rather efficiently with the cold Fermi
sea (the energy dependence of electron-electron scattering will be discussed in section \ref{sec-theory}), and thereby
excite electron-hole pairs (in this case, ``hole'' means a missing electron in the Fermi sea, not a valence band hole).
If the collector barrier has a suitable height, as in the center of Fig.\ \ref{fig-sample}(b), it will separate
excited electrons from the Fermi sea holes. While the electrons pass the barrier, the positively charged holes are
trapped between BE and BC. Without a connection to the environment, a positive charge would accumulate
here\cite{hotelectrons}, but since the side contact is grounded and therefore provides a reservoir of charge carriers,
electrons are drawn from this contact into the device. The jet pump analogy is therefore especially
appealing as it incorporates the attractive force exerted on the ``fluid`` in the side port. 

\section{Influence of the collector barrier}

\subsection{Calibration of collector barrier height}
\label{sec-calibration}

The collector barrier BC is first and foremost characterized by the applied gate voltage, \vBC, but its height, \eBC,
compared to the Fermi energy would be more useful. We have determined the actual
height of a barrier in units of energy (for barriers below the Fermi energy) by measuring
the reflection of Landau levels at the barrier in a perpendicular magnetic field\cite{komiyama,haug} as in Refs.
\onlinecite{hotelectrons} and \onlinecite{georg}.

\begin{figure}
\includegraphics[width=\columnwidth]{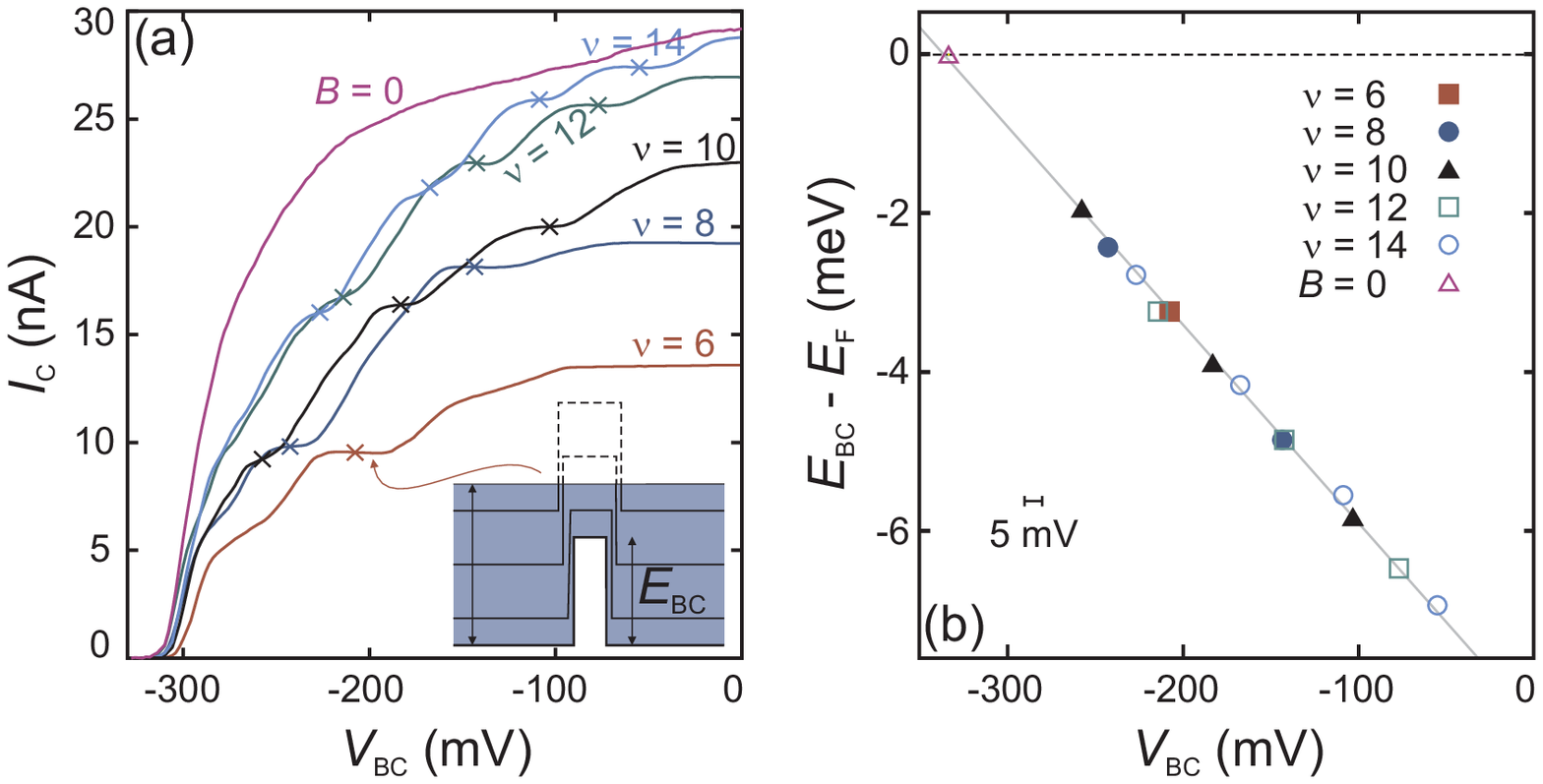}%
\caption{\label{fig-calib}(Color online) (a) Pinch-off curves of barrier BC in a perpendicular
magnetic field at integer filling factors
$\nu$. Reflection of landau levels at BC creates plateaus in the curves (see
sketch); crosses mark data points used for the calibration. (b) Points: allocated energies as a
function of gate voltages at plateau centers extracted from the set of curves
shown in (a) and corresponding energies, ``$B = 0$'' denotes additional
calibration for zero field (see text); line: fit of all data points, used for determining conversion
between \vBC\ and \eBC.}
\end{figure}

In contrast
to the experiments described in the rest of the article, these calibration measurements are
performed in the linear-response regime using the lock-in technique with $V_{E,\text{rms}} =
75\;\mu\text{V}$ at \unit{18.4}{\hertz} ($V_{E,\text{rms}}$ is kept small to minimize distortion
of the barrier
shape due to a voltage drop across the barrier). Fig. \ref{fig-calib}(a) 
plots the ac collector current, \iC, in a two-terminal measurement (side contact floating) as a
function of the voltage, \vBC, which controls the barrier height \eBC. Pinch-off curves for different magnetic fields
with integer bulk filling factors $6 \le \nu \le 14$ in the undisturbed 2DES are shown.

The inset of Fig.\ \ref{fig-calib}(a) demonstrates how the reflection of
Landau levels can be used in this setup to
extract information about the barrier height (sketch for filling factor $\nu =
6$): At the position of the barrier, the
number of occupied Landau levels is reduced. The higher the barrier, the more
Landau levels are pushed above the Fermi
edge and therefore do not contribute to the transmission. As long as the number
of Landau levels between the top of the
barrier and the Fermi energy does not change, the transmission should stay
constant, and a plateau in the current is
expected. At the center of the plateau we have $\eF - \eBC = k \; \hbar
\omega_c$ with $k \in {1, 2, \dots, \nu/2}$.
The plateau positions in \vBC, and the respective value of $k$, can be determined for several bulk filling factors
$\nu$ as shown
in Fig.\ \ref{fig-calib}(b). We estimate the error of the plateau position to be
about
\unit{5}{\milli\volt} [as marked in Fig.\ \ref{fig-calib}(b)]. The energy values are much more accurate since their
main error
source is an inaccuracy in the magnetic field value, e.\ g., due to
ferromagnetic material. Shubnikov-de Haas oscillations periodic in $1/B$ observed in the same measurement run
suggest a neglibile error in $B$ and therefore in energy. The pinch-off curve for $B = 0$ yields one additional
data point, the gate voltage corresponding to $\eBC = \eF$ [marked by ``$B = 0$`` in Fig.\ \ref{fig-calib}(b)] at
which current starts to flow across the barrier in a two-terminal setup. A linear fit to all datapoints yields the
relation $\eBC =-0.025 \;e \vBC - \unit{8.4}{\milli\evolt}$ as our final barrier calibration.

The barriers used in the experiments presented here turned out to be
sufficiently stable over a long period of time so
that it was enough to perform the calibration once per barrier. The only
exception was a sudden dramatic shift of the pinch-off
curves of a single barrier (in the order of \unit{300}{\milli\volt} toward more positive
voltages). Those changes were
irreversible, seemingly not caused by external influences, and only happened once per barrier. Since they were easy
to detect, they did not consitute a serious problem, only the calibration had
to be repeated. The measurements shown in Figs.\ \ref{fig-sample}(b),  \ref{fig-eyes}, and \ref{fig-stripes} have
been performed after the barrier had changed, hence $\vBC > 0$. For this set of data, the calibration relation, $\eBC
=-0.026 \;e \vBC - \unit{0.35}{\milli\evolt}$ was obtained.

\subsection{Tuning for amplification}

\begin{figure}
\includegraphics[width=\columnwidth]{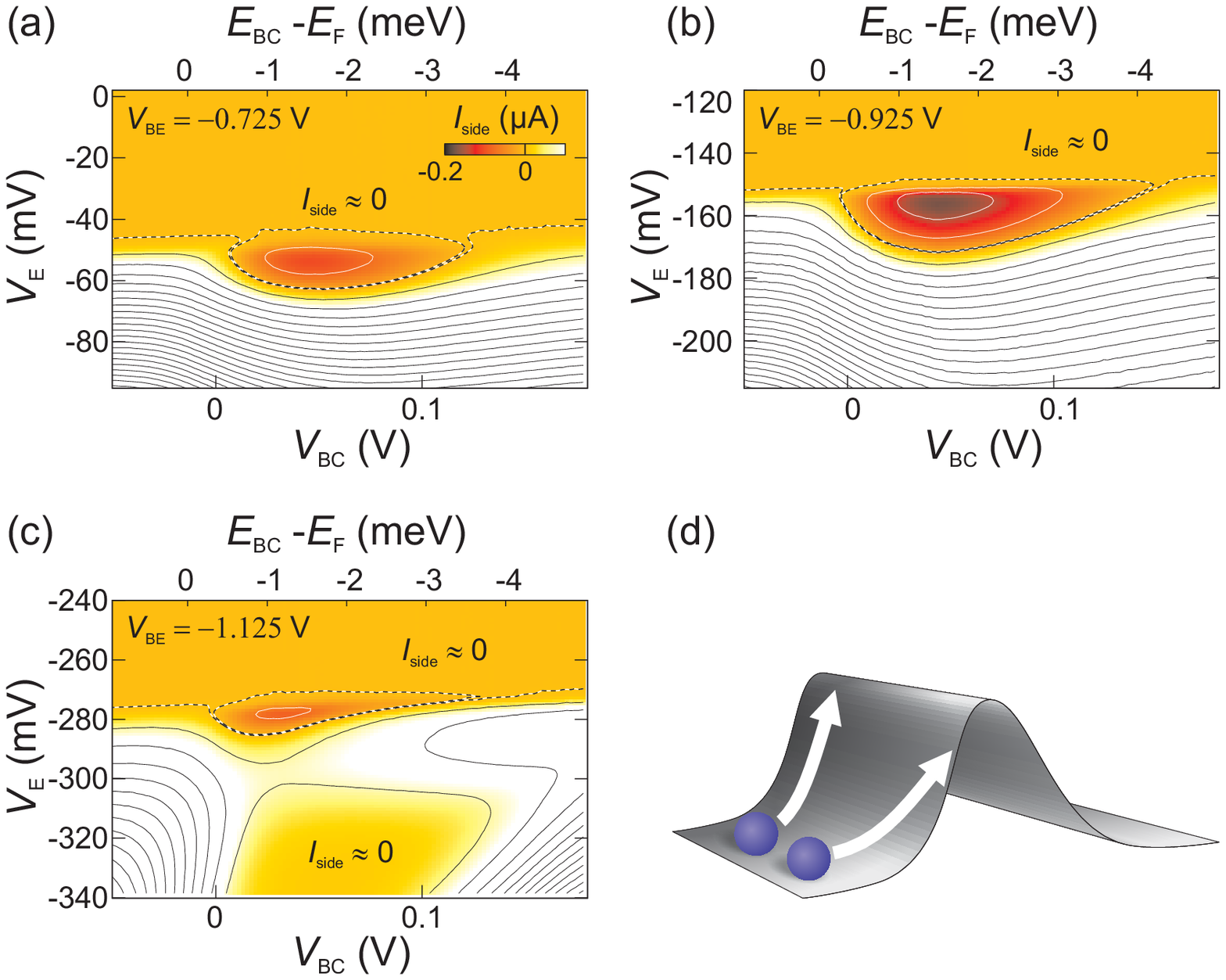}
\caption{\label{fig-eyes}(Color online) Side current as a function of collector barrier voltage
\vBC\ and bias voltage \vE. Collector barrier height calculated from \vBC\ as shown in section
\ref{sec-calibration} is depicted on upper axis. Contour lines spaced by \unit{70}{\nano\ampere} are drawn in black
for $\iside >0$, in white for $\iside < 0$, and dashed lines for $\iside \approx 0$. Emitter barrier voltage \vBE\ is
(a) \unit{-0.725}{\volt}, (b) \unit{-0.925}{\volt}, and (c)
\unit{-1.125}{\volt}; (d) sketch to demonstrate 2D model of barrier
height influence (see main text for details).
}
\end{figure}

Fig.\ \ref{fig-eyes}(a)--(c) 
show measurements of \iside\ as a function of collector barrier height (on the top axis;
the corresponding gate voltage \vBC\ is shown on the bottom axis) and bias voltage \vE. In the upper part of the graphs,
$\iside \approx 0$, since here the emitter is closed. The current starts to flow into the device at a threshold bias,
e.\ g.,
$\vEt \approx \unit{-150}{\milli\volt}$ for Fig.\ \ref{fig-eyes}(b). Upon crossing the threshold \iside\ immediately
becomes negative in the central area of the plots (framed by a dashed line marking $\iside = 0$), corresponding to
amplification. For larger bias voltages, the side current changes sign and
quickly increases ($\iside < 0$). The latter effect is actually related to an increase in the total current flowing
through the device and has been discussed in detail in Ref.\ \onlinecite{hotelectrons}.

From Figs.\ \ref{fig-eyes}(a) to \ref{fig-eyes}(c), \vBE\ is made more negative, which has several implications. One
consequence is a shift in the
threshold bias \vEt\ to larger energies since the emitter is more closed for more negative \vBE. In addition, the
area of $\iside < 0$ and the magnitude of \iside\ depend on $\vEt (\vBE)$, with the largest effect visible in
Fig.\ \ref{fig-eyes}(b). More details, including a discussion of the area showing $\iside \approx 0$ at large
\vE\ [Fig.\ \ref{fig-eyes}(c)], will be given in section \ref{sec-theory}.

\subsection{Model}
Fig.\ \ref{fig-eyes} demonstrates that the electron jet pump behavior depends strongly upon the
collector barrier height. Strikingly, $\iside < 0 $ is exclusively found when BC is below the Fermi energy
($\eBC < \eF$). This excludes heating as the reason of the observed effect since in this case the maximum effect would
be expected for $\eBC > \eF$. In a na\"{i}ve one-dimensional
model based on non-equilibrium electron-electron scattering (Sec. \ref{sec-jetpump}), the BC exactly at the Fermi energy
would result in the best charge
separation since then all excited electrons (above \eF) would pass the barrier while all holes (below \eF) would be
reflected. Maximal amplification would therefore be expected at $\eBC = \eF$, and the area of $\iside < 0 $ would
roughly be centered around this point.

The device studied here is two-dimensional (2D) in nature, and in 2D the very simple model has to
be modified. In 1D, it was sufficient to look at the total kinetic energy of an electron to determine whether it
will pass the barrier or it will be reflected. In 2D, only the forward momentum component, $p_\perp$, perpendicular to
the
barrier is significant. A charge carrier can only cross the barrier if  $p_\perp^2/2m > \eBC$ is fulfilled, thus passing
the
barrier is harder for particles not perpendicularly hitting it. A simple classical analogy to this situations is
depicted in Fig.\ \ref{fig-eyes}(d), showing two balls rolling towards a hill with the same velocity but at different
angles. The ball hitting the barrier perpendicularly will pass more easily than the one moving at an angle. If one now
considers a large amount of charge carriers with a distribution of angles in 2D, less carriers will
cross a barrier of the same height as compared to the 1D case. In other words, the barrier has to be lowered, compared
to 1D, to reach a comparable
amount of passing charge carriers. This explains why the jet pump effect is shifted to lower barrier heights ($\eBC <
\eF$) than predicted by the simple 1D model.

\section{Electron-electron scattering length}
\label{sec-theory}

\begin{figure}
\includegraphics[width=\columnwidth]{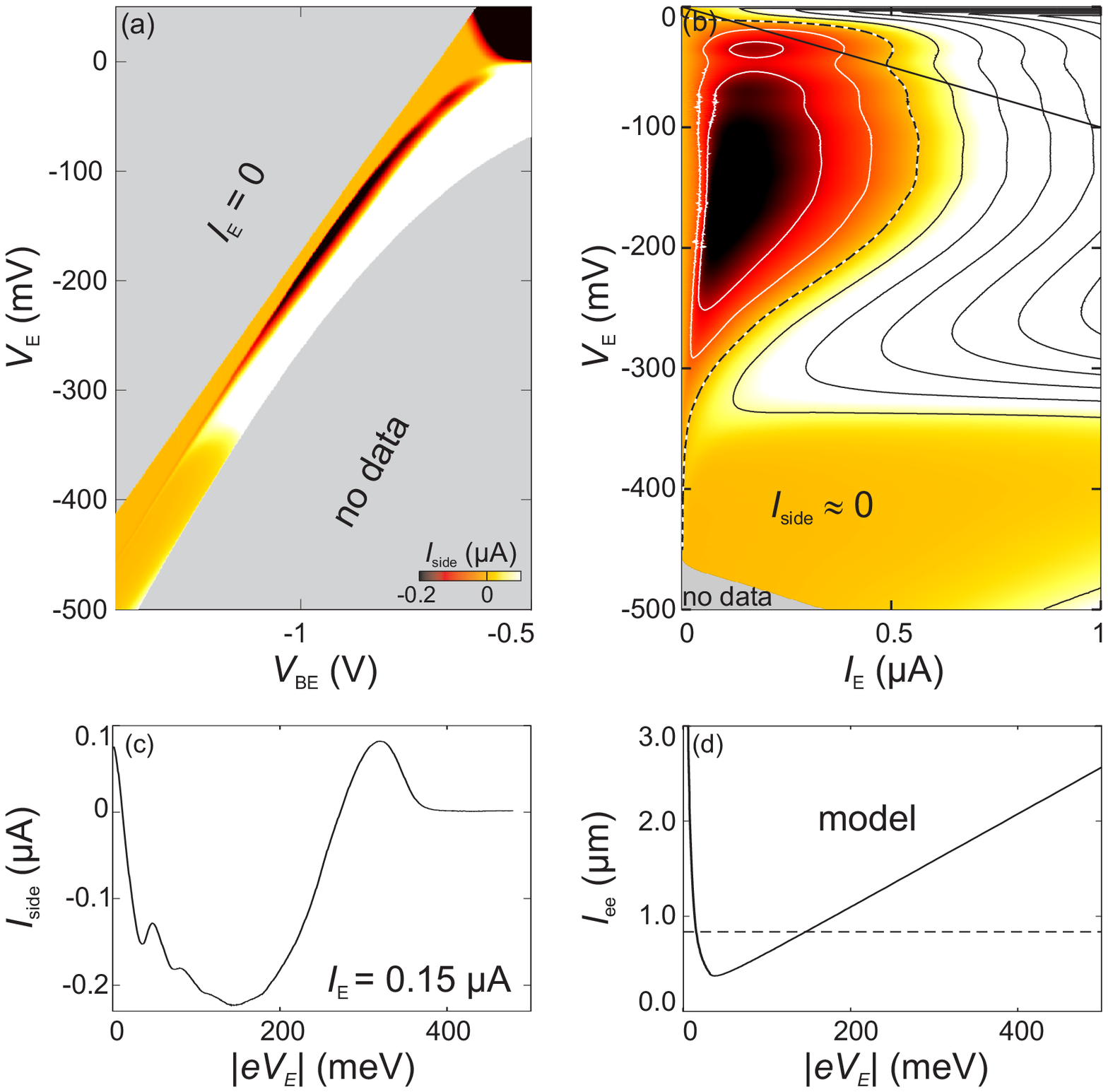}
\caption{\label{fig-stripes}(Color online) (a) \iside\ as a function of \vBE\ and \vE\ measured for dissipated
powers $\left| \vE\times\iE \right| \le700\,$nW. No data exist for higher powers (lower right corner) and for the upper
left
corner; here the emitter QPC is closed and all currents vanish. (b) Same data as in (a), plotted as a function of
injected current, \iE; contour lines are spaced by \unit{70}{\nano\ampere}, $\iside \approx 0$ is marked by a dashed
line. (c)
Vertical slice of Fig.\,\ref{fig-stripes}(a)
at $\iE = 0.15\;\mu\text{A}$;
(d) numerical calculations of electron-electron scattering length, \lee, as a
function of excess kinetic energy $|e\vE| \simeq E_{\text{kin}}-\eF$ at $T=0$; the dashed line marks sample dimensions.
}
\end{figure}

In the \iside\ measurements presented up to now, the collector barrier (\vBC) was varied while the emitter barrier
(\vBE) was kept constant. It is also instructive to analyze data for a fixed \vBC\ while \vBE\ is varied. An example of
such a
measurement is shown in Fig.\ \ref{fig-stripes}(a).
The threshold of nonvanishing current through the device is visible along a roughly diagonal line. Above that, in the
upper left corner, all currents are zero; therefore, most of this area has not been mapped out in detail. The lower
right corner also contains no measured data points, since here, at rather open emitter and large negative bias, the
power dissipated in the device would be very high. For the actual measurement, power was therefore limited to
$\left| \vE\times\iE\right| \le700\,$nW.

In an approximately diagonal stripe tapered at both ends, $\iside < 0$ is visible (in addition, in the upper right
corner a region with $\iside < 0$ due to ohmic behavior is observed at $\vE > 0$). The data show the same general
behavior already visible in Figs.\ \ref{fig-eyes}(a)--\ref{fig-eyes}(c).
It is far easier to analyze another representation of the data, depicted in Fig.\ \ref{fig-stripes}(b), which shows
\iside\
as a function of \vE\ and the total current $\iE = \iC + \iside$ (\iside\ and \iC\ were measured).  Below the straight
solid
line the resistance of the emitter is $\left| \vE \right| / \iE > \unit{100}{\kilo\ohm}$ (contact resistances are much
smaller). The emitter is thus almost pinched off, and we can assume that all electrons contributing to \iE\ are
injected at BE with an energy close to $\left| e \vE \right|$. Vertical (horizontal) slices of Fig.\
\ref{fig-stripes}(b) therefore show \iside\ as a function of energy (power) at constant \iE\ (energy per
electron) (see Ref.\ \onlinecite{hotelectrons}). Here we concentrate on the energy dependence.

Fig.\ \ref{fig-stripes}(c) shows a slice of Fig.\ \ref{fig-stripes}(b) at constant total current, allowing one to
analyze the dependence of \iside\ upon excess kinetic energy $\left| e \vE \right|$ right at the maximum of the observed
effect (most negative \iside). For very small $\left|\vE \right|$, \iside\ is positive, then rapidly decreases to
reach
its minimum value at an energy of $\left| e \vE \right| \approx \unit{150}{\milli\evolt}$. For larger energies \iside\
again increases and takes positive values.  However, for $\left| e \vE \right| > \unit{300}{\milli\evolt}$ \iside\
decreases once more, and then vanished in the high-energy limit.  The latter phenomenon is also visible in Fig.\
\ref{fig-stripes}(b) as an extended area of $\iside \approx 0$ as well as in Fig.\ \ref{fig-eyes}(c). 

The behavior of \iside\ as a function of $\left| e \vE \right|$ is closely related to the energy dependence of the
electron-electron scattering length, \lee. Predictions of \lee\ near the linear response regime have been
made before,\cite{chaplik, giuliani} but to describe scattering of a single electron with a 2DES, at a kinetic
energy greatly exceeding \eF, an extension of
those earlier models is necessary. We have performed numerical calculations for $T = 0$ based on the random phase
approximation to determine \lee\ as a function of excess kinetic energy for the whole energy range accessible in the
experiments presented here. The result is shown in Fig.\ \ref{fig-stripes}(d). 
As the kinetic energy $\ekin = \left|e \vE \right| + \eF$ exceeds \eF, electron-hole excitations cause a rapid
decrease of \lee\ as a function of $\left|e \vE \right|$ [$\lee \propto  1/((p-p_{\text{F}})
\ln(\left|p-p_{\text{F}}\right|))$]. The subsequent increase of $\lee \propto \left|e \vE \right|$ toward high kinetic
energies ($\ekin \gg \eF$) is caused by a decreased interaction time in combination with a suppressed plasmon radiation.
This result compairs fairly well with its three-dimensional (3D) counterpart\cite{pines}. A major reason for this
similarity is that plasmon radiation in 3D is also suppressed below a threshold energy, although with a different
origin compared to 2D\cite{giuliani}.

The behavior of \lee\ can be mapped onto the measured energy dependence of \iside\ [Fig.\
\ref{fig-stripes}(c)] if the sample geometry is taken into account. A dashed horizontal line in Fig.\
\ref{fig-stripes}(d) marks \unit{840}{\nano\meter}, the distance between BE and BC. Electrons injected with energies
corresponding to a \lee\ smaller than this distance have a high probability of scattering between BE and BC, thereby
contributing to the jet pump effect by creating electron-hole pairs in the central region. Energies corresponding to a
small \lee\ and a positive slope of the curve in Fig.\ \ref{fig-stripes}(d) are even more favorable since hot electrons
always lose energy in scattering with the Fermi sea, thus after one scattering event the scattering length can be
reduced even further. This is likely to result in multiple scattering processes which produce many electron-hole pairs,
leading to a very negative \iside. As $\left| \vE \right|$ is increased further, \lee\ exceeds the
sample dimensions, and scattering events
tend to happen beyond BC. In an intermediate regime, scattering beyond BC, but still close to the barrier, may lead to
scattered electrons traveling back accross BC and
into the side contact which causes a positive \iside, which is visible in Fig.\ \ref{fig-stripes}(c) as a local maximum
at around
\unit{320}{\milli\evolt}. At the
highest energies studied here, $\iside \approx 0$, which is consistent with the very large value of \lee\ predicted by
our numerics. Here, electrons move ballistically through the sample and scatter only very far away from BC so that no
electron-hole separation occurs. No charge carriers reach the side contact, and $\iside = 0$.

\section{Influence of magnetic field}

\begin{figure}
\includegraphics[width=\columnwidth]{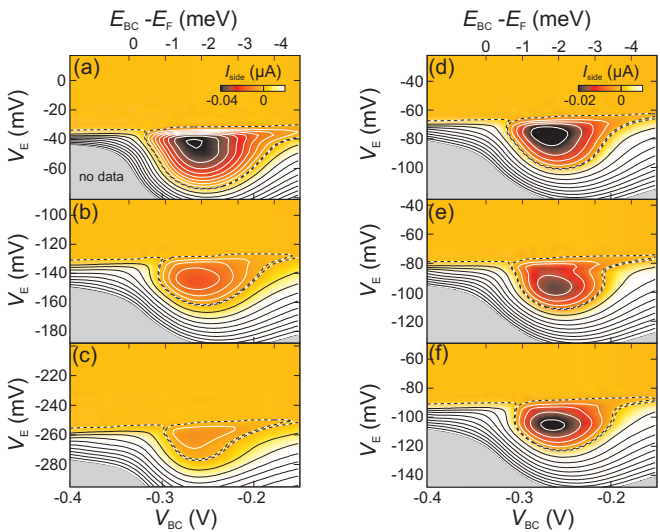}
\caption{\label{fig-magneticeyes}(Color online) Measurements similar to those in Fig.\ \ref{fig-eyes} with a magnetic
field of
\unit{5.2}{\tesla} perpendicularly applied. Contour lines spaced by \unit{5}{\nano\ampere} for $\iside < 0$ (white) and
\unit{10}{\nano\ampere} for $\iside > 0$ (black). Emitter barrier voltage $\vBE =
\unit{-0.675}{\volt}$ in (a), \unit{-0.875}{\volt} in (b), \unit{-1.075}{\volt}
in (c), \unit{-0.750}{\volt} in (d), \unit{-0.775}{\volt} in (e), and
\unit{-0.800}{\volt} in (f).
}
\end{figure}

Scattering lengths are expected to change considerably if external parameters are varied. Here the
influence of a magnetic field perpendicular to the two-dimensional electron system ist studied. Figures
\ref{fig-magneticeyes}(a)--\ref{fig-magneticeyes}(c)
show measurements similar to those presented in Fig.\ \ref{fig-eyes}(a)--(c), with an additional perpendicular
magnetic field of $B = \unit{5.2}{\tesla}$. The field direction is ''upwards,'' i.\ e., electrons injected into the
central part of the sample are guided to their left, away from the side contact. Data with and
without the magnetic field look rather
similar. However, the magnitude of the negative
side current is smaller by roughly a factor of 5 (note different color scale compared to Fig.\ \ref{fig-eyes}) while
the overall current passing through the device is virtually unchanged. A regime of $\iside \approx 0$
has been observed at high energies as in the case of $B = 0$, but it is not included in the set of data shown here.

Figures \ref{fig-magneticeyes}(d)--\ref{fig-magneticeyes}(f) show a series of measurements at more closely spaced
emitter barrier voltages
of $\vBE = \unit{-0.750}{\volt}$ in (d), \unit{-0.775}{\volt} in (e), and
\unit{-0.800}{\volt} in (f). The color scale is different from Fig.\
\ref{fig-magneticeyes}(a)--\ref{fig-magneticeyes}(c) to show the detailed
structure of the data. Here a nonmonotonic dependence on \vBE\ not visible in the overview
series shown in Figs.\ \ref{fig-magneticeyes}(a)--\ref{fig-magneticeyes}(c) is observed. Here, \iside\ is less negative
in Fig.\ \ref{fig-magneticeyes}(e) compared to Fig.\ \ref{fig-magneticeyes}(d) and \ref{fig-magneticeyes}(f), and
shows a
peculiar structure inside the area of $\iside < 0 $: two minima with a
lighter stripe in between. These substructures are related to the emission of optical phonons which lead to a periodic
reduction of negative side current as a function of kinetic energy; the period being \unit{36}{\milli\evolt}, which
is the
energy of optical phonons in GaAs\cite{hickmott_magnetophonon}. Traces of optical phonon
emission are already visible in the zero-field data presented in Fig.\ \ref{fig-stripes}(b) and \ref{fig-stripes}(c) at
low energies as
oscillations of $\iside(\vE)$. The emission of optical phonons and its relation to the electron jet pump is discussed
in detail in Ref.\todo{onlinecite}

\section{Conclusion}

We have studied the electronic Venturi effect in a relatively simple device containing three current-carrying contacts
and two barriers. Here the influence of the second, ``collector,'' barrier has been investigated in detail, since it is
vitally
important to create an electron jet pump. Such a device might have an application in amplifying small currents or
charges down to single electrons.

\vspace{-2mm}
\begin{acknowledgments}

We thank J.\,P.\ Kotthaus and S.\ Kehrein for fruitful
discussions. Financial support by the German Science Foundation via Grant Nos. SFB 631, SFB 689, LU 819/4-1, and the
German
Israel program DIP, the German Excellence Initiative via the "Nanosystems Initiative Munich (NIM)", and LMUinnovativ
(FuNS) is gratefully acknowledged. 
\end{acknowledgments}

%

\end{document}